\documentstyle[a4wide,rotating,epsfig,11pt]{article}
\newcommand{\bildchen}[3]{%%%%%%%%%%%%%%%%%%%%%%%%%%%%%%%%%%%
\begin{center}                                              %
\begin{sideways}                                            %
\makebox{\hspace*{5ex} \Large $\displaystyle #2$}           %
\end{sideways}                                              %
\mbox{{\epsfig{figure=#1,width=13.cm,%                      %
bbllx=1.8cm,bblly=9.2cm,bburx=20.cm,bbury=19.cm}}}          %
\end{center}                                                %
\begin{flushright}                                          %
   {\Large $\displaystyle #3$ \hspace*{5ex}}                %
\end{flushright}}                                           %
\begin{document}

\makebox[14cm][r]{May 1995}\par
\makebox[14cm][r]{hep-ph/9505434}\par
\vspace{.7cm}
\centerline{\Large \bf
RADIATIVE CORRECTIONS TO $\pi_{l2}$ AND $K_{l2}$ DECAYS}
\par
\vspace{1.cm}
\centerline{\sc Markus Finkemeier\footnote{%
Address from July 1995: Lyman Laboratory of
Physics, Harvard University, Cambridge MA 02138, USA}
}
\par \vspace*{0.5cm}
\centerline{INFN - Laboratori Nazionali di Frascati} \par
\centerline{P.O. Box 13}\par
\centerline{00044 Frascati (Roma), Italy}\par
\par
\normalsize

\begin{abstract}
We report on our reexamination of the radiative corrections to $\pi_{l2}$
and $K_{l2}$ decays.
We perform a matching calculation,
using a specific model with vector meson
dominance
in the long distance part and the
parton model in the short distance part.
By considering the dependence on the matching scale and on the hadronic
parameters, and by
comparing with model independent estimates, we
scrutinize the model dependence of the results.
For the pseudoscalar meson decay constants, we extract the values
$f_\pi = (92.1 \pm 0.3)\,\mbox{MeV}$ and $f_K = (112.4
\pm
0.9) \, \mbox{MeV}$. For the ratios $R_\pi$ and $R_K$ of the electronic
and muonic decay modes, we predict $R_\pi = (1.2354 \pm 0.0002) \cdot
10^{-4}$ and $R_K = (2.472 \pm 0.001) \cdot 10^{-5}$.
\end{abstract}
\begin{center}{ \em
Invited talk at the \\
Second Workshop on Physics and Detectors for Daphne\\
INFN - LNF\\
Frascati, April 4--7, 1995
}
\end{center}
\section{-- Introduction}
\subsection{\underline{What is interesting about $\pi_{l2}$, $K_{l2}$ decays?}}
Interest in $\pi_{l2}$ and $K_{l2}$ comes from two different sources.
On the one hand, the decay widths of the pion and the kaon into muon and
neutrino are used to extract the decay constants $f_\pi$ and $f_K$,
\begin{eqnarray*}
   \Gamma(\pi\to\mu\nu_\mu) & \longrightarrow &f_\pi
\\
   \Gamma(K\to\mu\nu_\mu)& \longrightarrow & f_K
\end{eqnarray*}
which are important input parameters for chiral perturbation theory. And
so the question arises how $O(\alpha)$ radiative corrections affect these
these parameters.

On the other hand, the ratios $R_\pi$ and $R_K$ of the electronic and
muonic decay modes
\begin{equation}
   R_\pi = \frac{\Gamma(\pi\to e\nu_e)}{\Gamma(\pi\to \mu \nu_\mu)}
\qquad
   R_K = \frac{\Gamma(K\to e\nu_e)}{\Gamma(K\to \mu \nu_\mu)}
\end{equation}
can be predicted by theory. What makes $R_\pi$ and $R_K$ very
interesting is that hadronic uncertainties cancel to a very large extent
in these ratios and that the electronic decay channels are strongly
helicity suppressed in the standard model due to the $V-A$ structure of
the weak interaction.
Thus the ratios $R_\pi$ and $R_K$ are very sensitive to non
standard model effects (such as multi-Higges, leptoquarks or other%
\cite{Sha82}), which might induce an effective pseudoscalar hadronic weak
current.
And so $\pi_{l2}$ and $K_{l2}$ decays allow for low-energy precision
tests of the standard model, if we are able to understand the radiative
corrections $\delta R_{QED}$:
\begin{equation}
   R_\pi = \frac{m_e^2}{m_\mu^2} \frac{(m_\pi^2 - m_e^2)^2}
   {(m_\pi^2 - m_\mu^2)^2}
   \Big( 1 + \delta R_{QED} \Big)
   \Big( 1 + \delta R_{non-SM} \Big)
\end{equation}
The experimental precision tends to be better for $\pi_{l2}$
decays. However, non standard model effects $\delta R_{non-SM}$, might
be enhanced in $R_K$ by a factor of $m_K/m_\pi$, and therefore $K_{l2}$
are of comparable interest in testing the standard model.
\subsection{\underline{Current Experimental Situation}}
For $R_\pi$, there are recent results (from 1992/93), obtained by TRIUMF
and PSI\cite{triumf}, resulting in\cite{RPP94} a relative precision of
\begin{equation}
   \frac{\Delta R_\pi}{R_\pi} = 3 \times 10^{-3}
\end{equation}
The precsion of $R_K$ as quoted by the 1994 particle data book
\begin{equation}
   \frac{\Delta R_K}{R_K} = 4 \times 10^{-2}
\end{equation}
is one order of magnitude less, but one should note that this number
is based on 1972--76 data. It would be very interesting indeed to receive
some new experimental information on this important quantity $R_K$.
\subsection{\underline{Theory}}
There is a vast literature on the radiative corrections to $\pi_{l2}$
decays. % Let us mention some important contributions.
In 1958/59,
Berman and Kinoshita\cite{Ber58} calculated the radiative corrections to
$\pi_{l2}$ decays using a model with an effective pointlike pion.
There have been numerous later attempts to improve this simple model in
order to take into account hadronic structure effects.
Important milestones in the understanding of these effects were the 1973
paper by Terent'ev\cite{Ter73}, who derived a general theorem on the
leading hadronic structure dependent correction, and the 1976 paper by
Marciano and Sirlin\cite{Mar76}, who proved that the leading logarithm
of the radiative correction is not affected by the strong interaction,
if one considers the inclusive decay rate
$\Gamma(\pi\to\mu\nu_\mu) + \Gamma(\pi\to\mu\nu_\mu\gamma)$ integrated
over all photons.
In 1977, Goldman and Wilson\cite{Gol77} considered various models for
the strong interaction effects in the context of gauge theories of the
weak interaction.
In 1990, Holstein\cite{Hol90} was the first to point to the importance
of short distance radiative corrections in the extraction of $f_\pi$,
and in 1993, Marciano and Sirlin reconsidered $\pi_{l2}$ decays from a
modern point of view, collection all the leading model-independent
corrections and giving rough order of magnitude estimates for the
remaining model dependent corrections.

Last year,
we have performed an improved matching calculation\cite{Fin94} of
$\pi_{l2}$ and $K_{\l2}$ decays. In this calculation we include vector
mesons as explicit degrees of freedom (instead of matching simply the
point pion/kaon with the short distance part), calculate the full short
distance correction (instead of including only the leading logarithm)
and we scrutinize the size of the model-dependent corrections by
explicitely calculating them within a reasonable model (instead of just
roughly estimating their order of magnitude).
This calculation allows to obtain improved values both for the
pseudoscalar decay constants $f_\pi$ and $f_K$, and for the ratios
$R_\pi$ and $R_K$, with a clear controll over the (small) model-dependent
contributions.
\section{-- Outline of the Calculation }
%
%Let us explain this in a little more detail.
%
\subsection{\underline{General Considerations}}
To obtain the $O(\alpha)$ radiative corrections to $\pi\to\mu\nu_\mu$,
we have to
evaluate Feynman diagrams with one photon loop, where the photon can be
contracted twice, once or not at all to the leptonic side.
Because of the
infra-red divergences, we have to add to this the tree diagrams for
$\pi\to\mu\nu_\mu\gamma$, where the photon couples either to the
leptonic or to the hadronic side.
The essential task is then to understand the coupling of a (real or
virtual) photon to the hadrons.

In the treatment of the
strong interaction, one can distinguish three different regimes.
In the low energy regime, where momentum transfers are small compared
with a typical hadronic scale of about $1 \, \mbox{GeV}$, there are the
low energy theorems of QCD which give model independent predictions.
In the high energy regime, where the virtual photon momentum is large on
a hadronic scale, we can use the parton model and calculate the QED
corrections to the $l \nu_l u d$ operator.
In the intermediate energy regime of momentum transfers of the order of
$1 \, \mbox{GeV}$, however, there is no way to obtain model independent
predictions. This energy regime is dominated by the physics of vector
meson resonances and can be described by phenomenological models only.

And so in order to evaluate the loop diagrams, one separates the
integration over the loop momentum into a long distance part
$k_E^2 = 0 \cdots \mu_{cut}^2$ and the short distance part
$k_E^2 = \mu_{cut}^2 \cdots m_Z^2$, where $\mu_{cut}$ is a hadronic
scale of about $\mu_{cut} \approx 1 \, \mbox{GeV}$.

Then one can either choose to neglect the problem of the
vector meson energy regime and match the low energy regime and the short
distances directly at a scale of
about $\mu_{cut} = m_\rho^2$, as has been done by Holstein%
\cite{Hol90} and by Marciano and Sirlin\cite{Mar93}.
A matching scale of $m_\rho^2$, however, is somewhat too large for the
low energy regime, on the one hand, becaue for $P^2 \to m_\rho^2$, all
the neglected
higher orders $O(P^6)$, $O(P^8)$, \dots become large. On the other hand,
$m_\rho^2$ is somewhat too small for the short distance part, where,
after all, one is using asymptotically free quarks.

Therefore we used the alternative approach of including the vector
mesons as explicit degrees of freedom in the long distance part. This
allows to push up the matching scale up to $\mu_{cut} = 1 \cdots 2 \,
\mbox{GeV}$, rendering the short distance part more reliably.

However, this unavoidably introduces model-dependence. Thus we then
scrutinize the size of the model dependence in different ways. Firstly
we vary the matching scale from $\mu_{cut} = 0.75 \cdots 3 \,
\mbox{GeV}$ and the parameters of the hadronics within their
experimental uncertainties. Secondly, we determine the full size of the
contribution to the loop integral from intermediate scales ($0.5 \cdots
3 \, \mbox{GeV}$), which is the regime where the model dependence is
large. Thirdly, we compare our result with the leading model independent
contributions and then take the full size of the model dependent
contribution as a measure of the uncertainty.
\subsection{\underline{Long Distance Part}}
For the long distance part, we start from a model with an effective
point meson model, which should be a good approximation for very small
momentum transfers.
These point meson amplitudes are then modified along the lines of vector meson
dominance in order to account explicitely for the vector meson degrees of
freedom.
Consider, for example, the coupling $V(p,p')^\mu$
of the photon to a pion line.
In the point meson approximation, this coupling is given by scalar QED as
\begin{equation}
   V(p,p')^\mu_{point} = i e (p + p')^\mu
\end{equation}
where $p$ and $p'$ are the ingoing and outgoing pion momenta. But actually this
coupling {\em defines} the pion electromagnetic form factor $F_\pi$ via
\begin{equation}
   V(p,p')^\mu_{phys} = i e F_\pi (k^2) (p + p')^\mu
\end{equation}
where $V_{phys}^\mu$ denotes the physical photon-pion-pion coupling and $k = p'
- p$ is the photon momentum. $F_\pi(k^2)$ is known experimentally rather well
in the relevant regime of $k^2$ up to $1$ or $2 \, \mbox{GeV}$.
Thus in the point pion amplitudes, we replace $V(p,p')^\mu_{point}$ by
$V(p,p')^\mu_{phys}$, using the parametrization of $F_\pi$ determined in%
\cite{Kue90}. This modification determines in turn the appropriate modification
of the pion-photon-W boson seagull coupling, viz.\ by the requirement of gauge
invariance.
In the case of the kaon, the scalar QED point kaon-kaon-photon coupling has to
be replaced by a coherent superposition of $\rho$, $\omega$ and $\Phi$ vector
meson dominance.
In addition to these modified point meson diagrams, there are loops which
correspond to the so-called structure dependent (SD) radiation in the radiative
decays $\pi (K) \to l \nu_l \gamma$, where the emitted photon becomes
contracted with the leptonic side.
\subsection{\underline{Short Distance Corrections}}
For photons with large virtuality, $|k^2| \, {}^>_\sim \, (1 \dots 2 \,
\mbox{GeV})^2$, we can use the parton model.
Thus the first step is to
calculate the one-loop photonic corrections $\delta {\cal A}$
to
${\cal A}_0 = \frac{G_F}{\sqrt{2}} [ \bar{u}_\nu \gamma^\mu \gamma_- u_l]
[\bar{u}_d \gamma_\mu \gamma_- u_u]$.

Neglecting all masses except for $m_l$ and $\mu_{cut}$, we obtain the leading
logarithm of the radiative correction
\begin{equation}
   \delta {\cal A} = \frac{\alpha}{\pi} \frac{1}{m_l^2 - \mu_{cut}^2}
   \left( m_l^2 \ln \frac{m_Z}{m_l} -
   \mu_{cut}^2 \ln \frac{m_Z}{\mu_{cut}^2} \right) + \cdots
\end{equation}
where the dots indicate corrections which are not leading in the limit $m_Z^2
\to \infty$.

In the radiative corrections to the individual decay rates $M \to l \nu_\tau
(\gamma)$, this leading logarithm clearly dominates, and so for the extraction
of $f_\pi$ and $f_K$ its consideration is sufficient.
It depends, however, only very little on the lepton mass and thus cancels
almost completely in the ratios $R_\pi$ and $R_K$.
In view of the very high precision of the theoretical predicton for these
ratios, we therefore go beyond the leading logarithm and calculate the full
$\delta {\cal A}$.
This full result for $\delta {\cal A}$
is firstly no longer proportional to the Born amplitude
${\cal A}_0$
and secondly it depends on the relative momentum of the two quarks.
Therefore we firstly project onto the $J^P = 0^-$ component of the two quarks,
and secondly we integrate over the relative momentum $u p$ of the quarks in the
infinite momentum frame ($u = -1 \dots +1$). This leads to a result
\begin{equation}
   \Big( \delta R_M \Big)_{short\, dist} =
   \frac{3}{2 f_M} \int_{-1}^{+1} du \,  \Phi_M(u) \, r_M(u)
\end{equation}
Here $\Phi_M(u)$ denotes the parton distribution function, whereas $r_M(u)$ is
calculated from the projected short distance diagrams. We find that $r_\pi(u)$
and $r_K(u)$ depend only very little on $u$, and so we can approximate them by
their values at $u=0$, where the wave function is presumably peaked:
\begin{equation}
   \Big( \delta R_M \Big)_{short\, dist} \approx r_M(0)
   \underbrace{\frac{3}{2 f_M} \int_{-1}^{+1} du \,  \Phi_M(u) }_%
   {\displaystyle= 1}
    = r_M(0)
\end{equation}
where the last equation follows from a sum rule\cite{Bra70}.

In matter of fact, we find that $R_\pi$ and $R_K$ are also dominated by the
leading logarithm,
\begin{equation}
   \Big( \delta R_M \Big)_{short\, dist} \approx
   \frac{2 \alpha}{\pi} \frac{m_\mu^2}{m_\mu^2 - \mu_{cut}^2}
   \ln \frac{m_\mu}{\mu_{cut}}
\end{equation}
\subsection{\underline{Treatment of Real Photons}}
Because of the infra-red divergences, one has to add some or all real photons
to the virtual corrections and to consider inclusive decay rates
$$
    \Gamma(M \to l \nu_l) + \Gamma(M \to l \nu_l \gamma)
$$

Whether one includes only soft photons or all real radiation is a matter of
convention. Our convention, which appears to be in accord with the one used by
experimentalists, is to include all photons in the pion decays $\Gamma(\pi \to
l \nu_l (\gamma))$. In the case of the kaon decays $\Gamma(K\to l \nu_l
(\gamma))$, we include all the internal bremsstrahlung (IB) photons, but
exclude completely the structure dependent (SD) radiation.

For $K \to \mu \nu_\mu \gamma$ this SD radiation is completely negligible, but
for the electronic mode it is extremely large, $\Gamma_{SD}(K \to e \nu_e
\gamma) \approx \Gamma_0(K \to e \nu_e)$, where $\Gamma_0$ denotes the Born
amplitude.
Therefore it is usefull to consider the SD radiation as a separate decay mode
and not to include it in the radiative correction to $K \to e\nu_e$.

Of course it is not possible to tell definitely whether a radiated photon
is due to internal bremsstrahlung or to structure dependent radiation.
However, if suitable experimental cuts are used, which put a small upper
limit onto the photon energy,
the measured rate of $K \to
e \nu_e (\gamma)$ will include only a very small SD background,
and only very little of the IB part will have been discarded. Using
the predicted differential distributions\cite{Bro64,Bij92}, the SD
background can be subtracted and the missing IB part added. Because of
the smallness of this correction, it does not give rise to
any important uncertainties.

\section{ -- Numerical Results}
\subsection{\underline{The pseudoscalar decay constants}}
 Adding up the long and short distance corrections, we obtain the full
radiative correction. The numerical result depends on the choice of the
matching scale $\mu_{cut}$ and on the parameters of the hadronics, which are
known only with limited experimental precision.

%%%%%%%%%%%%%%%%%%%%%%%%%%%%%%%%% fig1
\begin{figure}
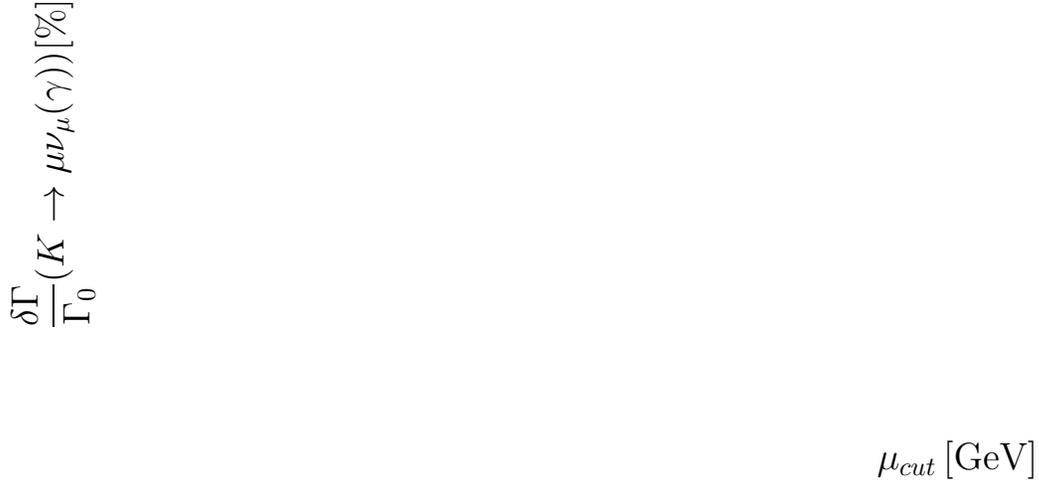

\caption{Radiative correction to $\Gamma(K\to\mu\nu_\mu)$, using
different choices for the hadronic parameters: Standard choice
(I, solid) and variations (II, dashed and III, dotted)}
\label{fig1}
\bildchen{fig2.ps}
{\frac{\delta \Gamma}{\Gamma_0}(K\to\mu\nu_\mu(\gamma)) [\%]}
{\mu_{cut}\,[\mbox{GeV}]}
\end{figure}
%%%%%%%%%%%%%%%%%%%%%%%%%%%%%%%%%%%%%%%%%%%%%%%%%%%%

In Fig.~\ref{fig1} we display the correction to the decay rate $\Gamma(K \to
\mu \nu_\mu(\gamma))$ in variation with $\mu_{cut}$, using three different
choices for the hadronic parameters. The solid line (I) corresponds to the
central values of the hadronic parameters, whereas the dashed (II) and the
dotted (III) lines are obtained by varying these parameters within reasonable
ranges.
{}From this, we obtain
\begin{equation}
   \frac{\delta \Gamma}{\Gamma_0} (K \to \mu \nu_\mu (\gamma))
   = ( 1.23 \pm 0.13 \pm 0.02) \% + O(\alpha^2) + O(\alpha \alpha_s)
\end{equation}
The central value $1.23 \%$ of the radiative correction of order $O(\alpha)$
has been obtained with $\mu_{cut} = 1.5 \,\mbox{GeV}$ and the central values of
the hadronics.
The first error quoted, $\pm 0.13 \%$ is the matching uncertainty, obtained
from varying the matching scale up and down by a factor of two, $\mu_{cut} =
0.75 \cdots 3 \, \mbox{GeV}$. The second error, $\pm 0.02 \%$ estimates the
uncertainty induced by the hadronic parameters.

In view of the smallness of the error, one has to worry about higher order
corrections. In\cite{Mar93}, the authors have summed up the leading
$O(\alpha^n)$ corrections for the leading logs in the short distance part,
which leads to an enhancement of $+0.13 \%$. Furthermore they considere the
leading QCD short distance corrections of order $O(\alpha \alpha_s)$, which
decrease the short distance part by $-0.03 \%$.
Similarly, $O(\alpha)^n$ effects should be considered in the long distance
part.

Furthermore, note that
we employ $m_Z$ as an ultra-violet cut-off for the short
distance corrections, according to the general theorems by Sirlin%
\cite{Sir78} on short-distance electroweak corrections to semileptonic
processes. But this implies that there is an uncertainty of the order
of $\alpha/(2 \pi) \times O(1) \approx 0.1\% $ in the definition of the
radiative
correction, because a change of the cut-off scheme would induce a change
of the result of this order. Note that the error of the $O(\alpha)$
correction which we have determined
is of the same order of magnitude as this inherent uncertainty.

Taking into account the higher order corrections and these uncertainties, we
use the following value in order to extract $f_K$:
\begin{equation}
   \frac{\delta \Gamma}{\Gamma_0} (K \to \mu \nu_\mu (\gamma))
   = ( 1.3 \pm 0.2) \%
\end{equation}
Using the 1994 particle data group values for the decay rates and the $V_{us}$
matrix element, this yields
\begin{equation}
   f_K = (112.4 \pm 0.9) \, \mbox{MeV}
\end{equation}
{}From a similar analysis, we obtain
\begin{equation}
   f_\pi = (92.1 \pm 0.1) \, \mbox{MeV}
\end{equation}
Two comments are in order:
\begin{itemize}
\item $f_\pi$ is not defined unambigous at the order $O(\alpha)$. One
could decide to absorb part of the radiative correction into $f_\pi$.
Here we have by definition factored out {\em all} radiative corrections from
$f_\pi$. Holstein\cite{Hol90} used a different definition which absorbs
terms proportional to $\ln(m_\pi/m_\mu)$ into $f_\pi$. Use of his
definition would yield a slightley lower value of $f_\pi = (91.9 \pm 0.1)
\, \mbox{MeV}$.
\item We have estimated the model dependence from the matching
uncertainty and the hadronic parameter dependence, and so the $\pm 0.1\,
\mbox{MeV}$ uncertainty is actually a lower limit.
\end{itemize}
Taking the full size of the model dependent part as uncertainty, a
conservative error estimate is
\begin{equation}
   f_\pi = (92.1 \pm 0.3) \, \mbox{MeV}
\end{equation}
Note that the error on $f_K$ is dominated by the error on $|V_{us}|$
rather than by model dependence.

Our number on $f_\pi$ is to be compared with the particle data group value%
\cite{RPP94}, which is based on\cite{Mar93}. Dividing by $\sqrt{2}$ in
order to conform with our convention, it reads
\begin{equation}
   f_\pi^{RPP} = (92.4 \pm 0.07 \pm 0.25) \, \mbox{MeV}
\end{equation}
where the first error $\pm 0.07$ comes from the experimental uncertainty
on $|V_{ud}|$ and the second $\pm 0.25$ from the matching uncertainty.
The numbers are compatible within the error bars. Our number has a
smaller central value due to the improved matching procedure (our
matching dependence is in fact an order of magnitude smaller.)
\subsection{\underline{The ratios $R_\pi$ and $R_K$}}
%
%%%%%%%%%%%%%%%%%%%%%%%%%%%%%%%%%%%%%% Fig.2
\begin{figure}
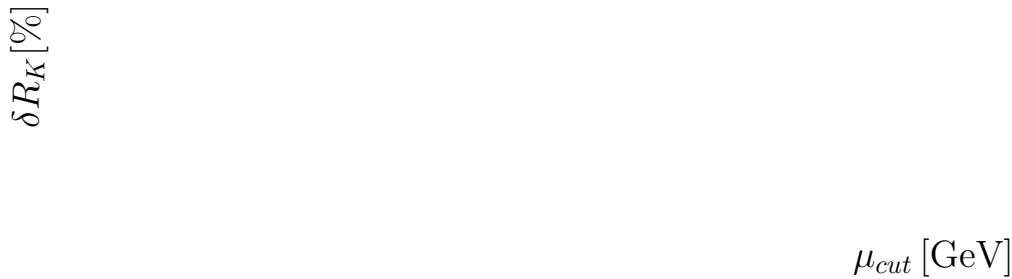

\caption{Radiative correction to the ratio $R_K $, using
different choices for the hadronic parameters: Standard choice
(I, solid) and variations (II, dashed and III, dotted)}
\label{fig2}
\bildchen{fig4.ps}
{\delta R_K [\%]}
{\mu_{cut}\,[\mbox{GeV}]}
\end{figure}
%%%%%%%%%%%%%%%%%%%%%%%%%%%%%%%%%%%%%%
Let us consider next the ratio $R_K$ of the electronic and muonic modes
in kaon decay.  From Fig.~\ref{fig2} we obtain
\begin{equation}
  \delta R_K = - (3.73 \pm 0.02 \pm 0.03) \% + O(\alpha^2)
\end{equation}
where the error given are the maching uncertainty $\pm 0.02 \%$ and the
uncertainty from the hadronic parameters $\pm 0.03\%$. These errors give
a lower limit of the model dependence of the result.

To get a reliable understanding of the model dependence, we have
compared our evaluation with the leading model-independent contributions
and added up the full sizes of the various model dependent parts. This
leads to an error estimate of $\pm 0.04 \%$.

Furthermore we have determined the contribution to the loop integral
from intermediate scales $\sqrt{|k^2| } = 0.5 \cdots 3 \, \mbox{GeV}$.
This also leads to an error estimate of $\pm 0.04 \%$.

Finally higher order corrections have to be discussed. In\cite{Mar93}
the authors sum up the leading logs in $m_e/m_\mu$ to all orders in $\alpha$,
leading to an
enhancement of $+ 0.05 \%$. Next-to-leading corrections are unknown,
and so we allow for an additional $\pm 0.01\%$ uncertainty.
This leads to our final result
\begin{eqnarray}
   \delta R_K & = & - (\underbrace{3.73 \pm 0.04}_{\displaystyle O(\alpha)}
   \quad  +
   \underbrace{0.05 \pm 0.01}_{\displaystyle O(\alpha^n), n \geq 2} ) \%
   = - (3.78 \pm 0.04) \%
\end{eqnarray}
and
\begin{equation}
   R_K = (2.472 \pm 0.001) \times 10^{-5}
\end{equation}
Similarly, we obtain
\begin{equation}
   R_\pi = (1.2354 \pm 0.0002) \times 10^{-4}
\end{equation}
\section{-- Summary}
Performing an improved matching calculation, we have extracted
\begin{eqnarray}
   f_\pi & = & ( 92.1 \pm 0.3 ) \, \mbox{MeV}
\nonumber \\
   f_K  & = & (112.4 \pm 0.9) \, \mbox{MeV}
\end{eqnarray}
where we have, by definition, factored out all $O(\alpha)$ effects from
$f_\pi$ and $f_K$.

Furthermore we found
\begin{eqnarray}
   R_\pi = \frac{\Gamma(\pi\to e \nu_e (\gamma))}
                {\Gamma(\pi\to \mu \nu_\mu (\gamma))}
   & = & (1.2354 \pm 0.0002) \times 10^{-4}
\nonumber \\
\nonumber \\
   R_K = \frac{\Gamma(K \to e \nu_e (\gamma))}
                {\Gamma(K \to \mu \nu_\mu (\gamma))}
   & = & (2.472 \pm 0.001) \times 10^{-5}
\end{eqnarray}
These are model independent predictions in the sense that the full model
dependence is included in the error bars given.

$R_\pi$ and $R_K$ can be predicted reliably within the standard model,
with relative uncertainties of a few $10^{-4}$. Thus they allow for very
sensitive low energy precision tests of the standard model.
\section*{Acknowledgements}
The author gratefully acknowledges financial support by the HCM program
under EEC contract number CHRX-CT90026
%


\begin{thebibliography}{99}
%
\bibitem{Sha82}
O. Shanker, Nucl. Phys. {\bf B204}, 375 (1982)
%
\bibitem{triumf}
%
D.I. Britton {et al.}, Phys. Rev. Lett. {\bf 68}, 3000 (1992);
C. Czapek {et al.}, Phys. Rev. Lett. {\bf 70}, 12 (1993)
%
\bibitem{RPP94}
Review of Particle Properties, L. Montanet {\em et al.}, Phys. Rev. {\bf
D50} 1173, (1994)
%
\bibitem{Ber58}
S. M. Berman, Phys. Rev. Lett. {\bf 1}, 468 (1958);
T. Kinoshita, Phys. Rev. Lett. {\bf 2},  477 (1959)
%
\bibitem{Ter73}
M.V. Terent'ev, Yad. Fiz. {\bf 18}, 870 (1973), Sov. J. Nucl. Phys.
{\bf 18}, 449 (1974)
%
%
\bibitem{Mar76}
W.J. Marciano, A. Sirlin, Phys. Rev. Lett. {\bf 36}, 1425 (1976)
%
\bibitem{Gol77}
T. Goldman and W.J. Wilson, Phys. Rev. {\bf D15}, 709 (1977)
%
\bibitem{Hol90}
B.R. Holstein, Phys. Lett. {\bf B244}, 83 (1990)
%
%
\bibitem{Fin94}
M. Finkemeier, ``Radiative Corrections to $\pi_{l2}$ and $K_{l2}$
decays'', LNF-94/076(IR), hep-ph/9501286
%
\bibitem{Mar93}
W.J. Marciano, A. Sirlin, Phys. Rev. Lett. {\bf 71}, 3629 (1993)
%
%
\bibitem{Kue90}
J. H. K\"uhn and A. Santamaria, Z. Phys. {\bf C48}, 445 (1990)
%
\bibitem{Bra70}
R. Brandt and G. Preparata, Phys. Rev. Letters {\bf 25}, 1530 (1970)
%
\bibitem{Bro64}
S. G. Brown, S. A. Bludman, Phys. Rev. {\bf 136}, B1160 (1964)
%
\bibitem{Bij92}
J. Bijnens, G. Ecker, J. Gasser, Nucl. Phys. {\bf B396}, 81 (1993)
%
\bibitem{Sir78}
A. Sirlin, Rev. Mod. Phys. {\bf 50}, 573 (1978)
%
%
\end{thebibliography}
\end{document}